\documentclass[12pt]{article}
\usepackage[pctex32]{graphics}
\begin{document}
\vspace{1cm}
\begin{center}
~\\
{\bf  \Large Holographic Description of Glueball and Baryon in  Noncommutative Dipole Gauge Theory}
\vspace{1cm}

                      Wung-Hong Huang\\
                       Department of Physics\\
                       National Cheng Kung University\\
                       Tainan, Taiwan\\

\end{center}
\vspace{1cm}
\begin{center}{\bf  \Large ABSTRACT } \end{center}
We study the glueball spectrum in the supersymmetric and non-supersymmetric 4D non-commutative dipole gauge theory from the  holographic description. We adopt the semiclassical WKB approximation to solve the dilaton and antisymmetric tensor field equations on the dual supergravity backgrounds to find the analytic formula of the spectrum of $0^{++}$ and $1^{--}$ glueballs, respectively.   In the supersymmetric theory we see that the dipole length plays the intrinsic scale which reflects the discrete spectrum therein.  In the non-supersymmetric theory,  the temperature (or the radius of compactification)  in there will now play the intrinsic scale and  we see that the dipole has an effect to produce attractive force between the gluons within the glueball.  We also study the confining force between the quarks within the baryon via strings that hang into the dipole deformed AdS geometry and see that the dipole could also produce an attractive force between the quarks.  In particular, we find that the baryon has two phases in which  a big baryon is dual to the static string while a small baryon is described by a moving dual  string .
\vspace{1cm}
\\
\\
\\
\begin{flushleft}
E-mail:  whhwung@mail.ncku.edu.tw\\
\end{flushleft}


\newpage
\section{Introduction}
The  holographics of AdS/CFT correspondence [1,2] provides a powerful method to investigate the strong coupling gauge theory in dual supergravitional description.  The  correspondence has been applied to investigate several problems in large $N_ c$ QCD such as the Wilson loop [3,4], the meson spectra/dynamics [5,6], baryon dynamics [7-9], glueball spectrum [2,10-14] and so on.  

In the original proposal [1] the 10D background is the $AdS_5\times O^5$.  It relates to the comformally supersymmetric gauge theory which does not exist any mass scale to describe the hadronic physics and does not show confinement.   Witten [2] was the first to suggest a reliable background which breaks both of the conformality and supersymmetry to describe the real physical world. In his description the $AdS$ space is replaced by the Schwarzschild geometry describing a black hole in the AdS space. It was found that the Witten's supergravity background  gives results that are in qualitative agreement with expectations for QCD at strong coupling. 

In the first path of this paper we will study the spectrum of $0^{++}$ and $1^{--}$ glueball in the 4D supersymmetric and non-supersymmetric non-commutative gauge field theory from the  holographic description.  The dual supergravity backgrounds are the near-horizon geometry of  extremal D3-branes and nonextremal D4-branes, after applying T-duality and  smeared twist, which show that a nonzero B field shall be with one leg along the brane worldvolume and other transverse to it.  The solutions are the dipole deformed $AdS_5\times O^5$ and dipole deformed Witten's supergravity background (AdS Schwarzschild spacetime ) respectively, which had been constructed in [15-17]. 

According to the holographics there is the correspondence between the chiral operators and the supergravity states [1].   For example, the operator $trF^2$ in four dimensions corresponds to the dilaton field of supergravity in ten dimensions. Therefore the scalar glueball $J^{PC} = 0^{++}$ in QCD which couples to $trF^2$  is related to the dilaton propagating in the supergravity background and its mass is computable by solving the dilaton wave equation [2,10]. In a similar way, the operator $d^{abc}F^a_{\mu\alpha}F^{b\alpha\beta}F^c_{\beta\nu}$, where $d^{abc}$ is the symmetric structure constant, will couple to the antisymmetric tensor field and the $1^{--}$ glueball spectrum could be found by solving the antisymmetric tensor field wave equation [10].

  In our analysis the semiclassical WKB approximation is adopted to solve the dilaton and antisymmetric tensor field equations on the supergravity background.  The analytic formula of the spectrum of $0^{++}$ and $1^{--}$ glueball are therefore obtained.   We see that, in the supersymmetric theory the dipole length will play the intrinsic scale which reflects the discrete spectrum in the bound states of glueball.  In the non-supersymmetric theory, however, the temperature (or the radius of compactification)  in there will now play the intrinsic scale and we therefore attempt to see how the dipole will modify the discrete glueball spectrum.   Our analysis show that the dipole has an effect to produce attractive force between the gluons within the glueball.

In the second path of this paper we will study baryon energy.  We adopt the  method in [8] to study the confining force between the quarks within the baryon via strings that ``hang" into the dipole deformed AdS geometry.  In the method of [8] the baron vertex was considered as the strings and fivebrane which are described in terms of separated actions.  The method was improved by Callen et.al. [9] in which the baryon is constructed from the vertex that is considered as the D5 brane wrapped on an $S^5$ on which N fundamental strings terminate and they are dissolved in it [7].   The method allows an unified description of fivebrane and strings and provides an explicit string theory representation of the baryon vertex.   The mathematics in Callen method is more involved and we will in this paper adopt  the method of [8] for simplicity.     

After the evaluation we have also seen that the dipole could produce an attractive force between the quarks within the baryon.  In particular, we find that baryon has a minimum radius when the dual string is a static configuration and it will transit into another phase which dual to the moving string configuration at short distant.

  In section II we follow [10] to analyze the glueball spectrum in  4D  supersymmetric and non-supersymmetric dipole gauge theory.  In section III we follow [8] to analyze the baryon energy. The last is devoted to a conclusion.

Note that in the non-commutative dipole field theory each field $\Phi_a$ is associated with a constant dipole length $\ell_a$ and we define the ``non-commutative dipole product" by $\Phi_a (x) * \Phi_b (x) = \Phi_a (x-\ell_b/2) ~\Phi_b (x +\ell_a/2).$  It is a nonlocal field theory and break Lorentz invariance.  As there is the supergravity solution which dual to the  non-commutative dipole field theory the physical particle may has nonzero dipole length therefore.  The phenomenal constrain on the value of dipole length has not yet been set down now.  Some properties of the non-commutative dipole field theory have been studied in [15-18].  The noncommutative dipole field theories are interesting by themselves and it has a chance of finding a CP violating theory [18]. It is also an appropriate candidate to study the interaction of a neutral particles with finite dipole moments, like neutrinos, with gauge particles like photons. There are some experimental evidences of such interactions, which cannot be described by the commutative version of the standard model of particles [18]. 
\section{Glueball in Dipole Field Theory}
\subsection{4D Supersymmetric Dipole Field Theory}
\subsubsection{Supergravity Background}
To find the supergravity solution dual to the noncommutative dipole theory we  start with the 10D (with coordinates  $t,x_1,x_2,x_3, w_a, a=1,...6$) type II supergravity solution describing N coincident near extremal D3-brane (with worldvolume coordinates  $t,x_1,x_2$).  Now, following the prescription in [15], we first apply the T-duality transformation on the $x_3$ axis. Then, considering the ``smeared twist" as we go around the circle of new $x_3$ axis (with radius $R_0$), i.e. the ``twisted" compactification will accompany a rotation between $w_1, ..., w_6$ by a matrix $M_{ab}$ in the following way :$ (t,x_1,x_2,x_3,w_a)\rightarrow (t,x_1,x_2,x_3+ 2\pi R_0,\sum_{b=1}^6 M_{ab}w_b)$, 
in which $M$ is an element of the Lie algebra $SO(6)$. After the smeared twist we finally  apply the T-duality on the $x_3$ axis.

   With a proper choice of  $M$ the dual supergravity solution used to describe the supersymmetric  noncommutative  dipole field theory is [15,16]
$$ds_{10}^2 = U^2\left(- dt^2+ dx_1^2+ dx_2^2+{ dx_3^2\over 1+B^2U^2\sin^2\theta_1\sin^2\theta_2}\right)\hspace{4cm}$$
$$+ {1\over U^2} \left(dU^2+ U^2d\Omega_5^2-U^4B^2\sin^4\theta_1\sin^4\theta_2 {a_3d\theta_3+a_4d\theta_4+a_5d\theta_5\over 1+U^2B^2\sin^4\theta_1\sin^4\theta_2}\right). \eqno{(2.1)}$$
$$e^{2\phi}= {1 \over  1+ U^2B^2\sin^4\theta_1\sin^4\theta_2},~~~
B_{3\theta_i}= - {a_i~U^2B\sin^4\theta_1\sin^4\theta_2 \over 1+U^2B^2\sin^4\theta_1\sin^4\theta_2 },\hspace{1.7cm}\eqno{(2.2)}$$
in which $a_3 \equiv \cos\theta_4 $, $a_4 \equiv - \sin\theta_3\cos\theta_3\sin\theta_4 $, and $a_5 \equiv \sin^2\theta_3\sin^2\theta_4$, where $\theta_i$ are the angular coordinates parameterizing the sphere $S^5$ transverse to the D3 brane.    The value $``B"$ in (2.1) is proportional to the dipole length $``\ell"$ defined in the ``non-commutative dipole product" in section I.
\subsubsection{Spectrum of $0^{++}$  Glueball}
Consider first the $0^{++}$  glueball mass.  In the supergravity description we have to solve the wave equation of dilation :
$$\partial_\mu\left(e^{-2\phi}\sqrt g~g^{\mu\nu}~\partial_\nu \Phi \right) = 0. \eqno{(2.3)}$$
We look for $\theta_i$-independent solution of the form
 $$\Phi= \rho(U) e^{ik\cdot x},~~~~~ k_\mu= \left({M\over \sqrt{1-\beta^2}},0,0,{M\beta\over \sqrt{1-\beta^2}}\right). \eqno{(2.4)}$$
The momentum $k_\mu$ is given by the Lorentz boost of the rest frame momentum $k_\mu = (M, 0, 0, 0)$. In other words, we consider the dilaton equation in the moving frame with the velocity $\beta$ in unit of the light velocity, as that in the Moyal noncommutative theory [13].   Note that the dual string  in a background with  NS-NS $B$ field is somewhat analogous to the situation when a charged particle enters a region with a magnetic field.  Thus, the string will be moving with  a velocity.  The necessary to consider the moving dual string was first found by Maldacena in investigating the Wilson loop in Moyal-type noncommutative theory [19].  Later, it is known that  the similar property also shows in the noncommutative dipole theory [16,17].

The equation for $\rho$ becomes
$$\partial_U\left(U^5 \sqrt{1+B^2U^2}~\partial_U \rho \right) + {M^2U \sqrt{1+B^2U^2}\over \sqrt{1-\beta^2}}\left(1-\beta^2 \left(1+B^2U^2\right) \right)  \rho= 0. \eqno{(2.5)}$$
To proceed, we let $U=\sqrt y$ and above equation becomes
$$\partial_y\left(y^3 \sqrt{1+B^2y}~\partial_y \rho \right) + {M^2 \sqrt{1+B^2y}\over 4\sqrt{1-\beta^2}}\left(1-\beta^2 \left(1+B^2y\right) \right)  \rho= 0. \eqno{(2.6)}$$
Next, we define $B y= e^z$ and  above equation becomes
$$\partial_z\left(e^{2z} \sqrt{1+e^{z}}~\partial_z \rho \right) + {M^2B^2e^{z}\over 4}\sqrt{1+e^{z}}~ \left(1-\gamma e^{z} \right) \rho= 0, \eqno{(2.7)}$$
in which $\gamma= \beta^2/\sqrt{1-\beta^2}$.  As we will solve above differential equation by the semiclassical WKB approximation we first define the wavefunction
$$\Psi(z) = \sqrt {f(z)} ~ \rho(z),~~~~~~~with~~~~~~f(z) = e^{2z} \sqrt{1+e^{z}}, \eqno{(2.8)}$$
then the wavefunction $\Psi(z)$ will satisfy the equation
$$ \Psi''(z) + V(z) \Psi(z)=0,~~~~V(z) = {M^2B^2 \left(e^{-z}-\gamma \right)\over 4} -{1\over2}{f''(z)\over f(z)}+{1\over4}\left({f'(z)\over f(z)}\right)^2. \eqno{(2.9)}$$
In the WKB approximation we know that
$$\left(n+{1\over2}\right) \pi= \int_{-\infty}^{z_0}dz~\sqrt {V(z)},\eqno{(2.10)}$$
in which $z_0$ is a  turning point determined as following. Consider the case with $M \gg1$ we have the approximation
$$V\approx   {M^2B^2\over 4} \left(\left(1+{3\over2 M^2B^2}\right) e^{-z}-\left(\gamma + {25\over4 M^2B^2} \right) \right).\eqno{(2.11)}$$
The two turning points are therefore at $z=-\infty$ and $z_0$ where
$$ z_0\approx \ell n\left({1+{3\over2 M^2B^2}\over \gamma + {25\over4 M^2B^2}}\right).\eqno{(2.12)}$$
The mass spectrum evaluated from the WKB approximation in the case of $M \gg 1$ after performing the integration in  (2.10) becomes
$$\left(n+{1\over2}\right) \pi \approx {MB \sqrt \gamma\over 2}\left(1+{25\over 8}{1\over M^2B^2\gamma}\right),\eqno{(2.13)}$$
which implies the following spectrum of the $0^{++}$  glueball:
$$M ={1\over 2B\sqrt \gamma} \left[{(2n+1)\pi }+\sqrt{(2n+1)^2\pi^2 -{25\over2}}~\right] .\eqno{(2.14)}$$
The dipole length $B$ and velocity factor $\gamma$ in there now play the intrinsic scale which reflects the discrete spectrum.  This property is like that in [13] in which Nakajima et. al had seen that the Moyal noncommutativity could introduce an intrinsic scale in glueball discrete spectrum.
 \subsubsection{Spectrum of  $1^{--}$  Glueball}
Consider next the $1^{--}$  glueball mass.  In the supergravity description we have to solve the wave equation of antisymmetric tensor field $A_{\mu\nu}$ :
$${3\over\sqrt g} \partial_{\mu_1}\left(\sqrt g~\partial_{[\mu_2}A_{\mu_3\mu_4]} ~g^{\mu_1\mu_2}~g^{\mu_3\mu}~g^{\mu_4\nu} \right)  -16 g^{\mu_3\mu}~g^{\mu_4\nu} A_{\mu_3\mu_4} = 0. \eqno{(2.15)}$$
As before we  look for $\theta_i$-independent solution of the form 
$$A_{\mu\nu} = \rho_{\mu\nu}(U) e^{ik\cdot x},~~~~~ k_\mu= \left({M\over \sqrt{1-\beta^2}},0,0,{M\beta\over \sqrt{1-\beta^2}}\right). \eqno{(2.16)}$$
In searching the $1^{--}$  glueball spectrum we let $\rho_{\mu\nu}(U) = \rho(U)~ \delta_{\mu1}\delta_{\nu2}$ [10] and the equation for $\rho$ becomes
$$\partial_U\left({U\over \sqrt{1+B^2U^2}}~\partial_U \rho \right) +{1\over U \sqrt{1+B^2U^2}}\left(\left({M^2\over U^2}\left(1-\gamma B^2U^2\right) \right)- {16\over3}\right) \rho = 0. \eqno{(2.17)}$$
To proceed, we first let $U=\sqrt y$ and then define $B y= e^z$, the above equation becomes
$$\partial_z\left({1\over \sqrt{1+e^{z}}}~\partial_z \rho \right) + {M^2B^2e^{-z}\over 4}{1\over \sqrt{1+e^{z}}}~ \left(1-\gamma~ e^{z} \right) \rho= 0. \eqno{(2.18)}$$
In the semiclassical WKB approximation the wavefunction $\Psi(z)$ defined by 
$$\Psi(z) = \sqrt {f(z)} ~ \rho(z),~~~~~~~with~~~~~~f(z) = {1\over \sqrt{1+e^{z}}}, \eqno{(2.19)}$$
will satisfy the equation
$$ \Psi''(z) + V(z) \Psi(z)=0,~~~~~~~~V(z) = {M^2B^2 \left(e^{-z}-\gamma \right)\over 4} -{5\over16} \left({e^z\over1+e^z}\right)^2 +{e^z\over 4(1+e^z)} -{4\over3}. \eqno{(2.20)}$$
In the case with $M \gg 1$ we have the approximation
$$V\approx   {M^2B^2\over 4} \left(\left(1+{3\over2 M^2B^2}\right) e^{-z}-\left(\gamma + {67\over 12 M^2B^2} \right) \right).\eqno{(2.21)}$$
The two turning points are therefore at $z=-\infty$ and $z_0$ where
$$ z_0\approx \ell n\left({1+{3\over2 M^2B^2}\over \gamma + {67\over 12 M^2B^2}}\right).\eqno{(2.22)}$$
As before, the mass spectrum evaluated from the WKB approximation in the case of $M \gg 1$  becomes
$$\left(n+{1\over2}\right) \approx {MB \sqrt \gamma\over 2}\left(1+{67\over 6}{1\over M^2B^2\gamma}\right),\eqno{(2.23)}$$
which implies the following spectrum of the $1^{--}$  glueball:
$$M ={1\over 2B\sqrt \gamma} \left[{(2n+1)\pi }+\sqrt{(2n+1)^2\pi^2 -{134\over3}}~\right] .\eqno{(2.24)}$$
The dipole length $B$ and velocity factor $\gamma$ in there also play the intrinsic scale which reflects the discrete spectrum. 

As the dipole length in the 4D supersymmetric dipole theory plays the intrinsic scale we could not see how it will affect the glueball spectrum we will in next subsection investigate the 4D non-supersymmetric dipole Theory.
\subsection {4D Non-supersymmetric Dipole Field Theory}
\subsubsection{Supergravity Background}
To consider the  non-supersymmetric 4D dipole theory at zero temperature we shall consider the supergravity background which is constructed form near-extremal D4-brane solutions, instead of D3-brane.  Follow the prescription of 2.1 we can find the proper background which is described by [17]
$$ds_{10}^2 = U^{3/2}\left[- \left( 1-{U_T^3\over U^3}\right)dt^2+ dw^2+ dx^2+ dy^2+{ dz^2\over 1+B^2U^2\sin^2\theta_1}\right]\hspace{4cm}$$
$$+ {1\over U^{3/2}} \left[\left( 1-{U_T^3\over U^3}\right)^{-1}dU^2+ U^2d\Omega_4^2-B^2 U^4\sin^4\theta_1 {(a_2d\theta_2+a_3d\theta_3+a_4d\theta_4)^2\over 1+B^2U^2\sin^2\theta_1}\right]. \eqno{(2.25)}$$
$$e^{2\Phi}= {U^{3/2} \over  1+ B^2U^2\sin^2\theta_1},~~~
B_{z\theta_i}= - {a_i~U^2B\sin^4\theta_1\over 1+B^2U^2\sin^2\theta_1},\hspace{1.7cm}\eqno{(2.26)}$$
in which $a_2 \equiv \cos\theta_3 $, $a_3 \equiv - \sin\theta_2\cos\theta_2\sin\theta_3 $, and $a_4 \equiv \sin^2\theta_2\sin^2\theta_3$, where $\theta_i$ are the angular coordinates parameterizing the sphere $S^4$ transverse to the D4 brane.  The temperature is related to $U_T$ by
$$T = {3\over 4\pi} \sqrt {U_T}.\eqno{(2.27)}$$
The value $``B"$ in (2.25) is proportional to the dipole length $``\ell"$ defined in the ``non-commutative dipole product" in section I. 
\subsubsection{Spectrum of $0^{++}$  Glueball}
To consider first the $0^{++}$  glueball mass in the supergravity description we have to solve the wave equation of dilation in (2.3). As before we look for $\theta_i$-independent solution of the form
 $$\Phi= \rho(U) e^{ik\cdot x},~~~~~ k_\mu= \left(k_0,0,0,0\right), \eqno{(2.28)}$$
and glueball mass $M^2 = k_0^2$.  The equation for $\rho$ becomes
$$\partial_U\left( \sqrt{1+B^2U^2} \left(U^3-U_T^3\right)~U\partial_U \rho \right) + \sqrt{1+B^2U^2}~U~ M^2 \rho= 0. \eqno{(2.29)}$$
To proceed, we let $U=\sqrt y$ and above equation becomes
$$\partial_y\left( \sqrt{1+B^2y}~\left(y^{3\over2}-y_T^{3\over2}\right)y~\partial_y \rho \right) + {M^2\over 4} \sqrt{1+B^2y}~\rho= 0. \eqno{(2.30)}$$
Next, we define $ y= y_T(1+e^z)$ and  above equation becomes
$$\partial_z\left( f(z)~\partial_z \rho \right) + {M^2e^{z}\over 4\sqrt {y_T}} \sqrt{1+B^2y\left(1+e^{z}\right)}~\rho= 0. \eqno{(2.31)}$$
where
 $$f(z) = \sqrt{1+B^2y_T\left(1+e^{z}\right)}\left(\left(1+e^{z}\right)^{3/2}-1\right)~e^{-z} \sqrt{1+e^{z}}.~\eqno{(2.32)}$$
In the  semiclassical WKB approximation we  define the wavefunction $\Psi(z) = \sqrt {f(z)} ~ \rho(z)$ and the wavefunction $\Psi$ will satisfy the equation
$$ \Psi''(z) + V(z) \Psi(z)=0,~~~~V(z) ={M^2\over 4\sqrt {y_T}}{e^{2z}\over \left(\left(1+e^{z}\right)^{3/2}-1\right)~\sqrt{1+e^{z}}} -{1\over2}{f''(z)\over f(z)}+{1\over4}\left({f'(z)\over f(z)}\right)^2. \eqno{(2.33)}$$
In the case of $M \gg 1$ the two turning points determined by the function $V(z)$ are at $z=-\infty$ and $z_0$ where
$$ z_0\approx 2~ \ell n\left[{16\lambda\over9} + \left({16\lambda\over9}\right)^3{B^2\over8}\right], ~~~~\lambda \equiv {M^2\over4\sqrt {y_T}}.\eqno{(2.34)}$$
As before, the mass spectrum evaluated from the WKB approximation becomes
$$\left(n+{1\over2}\right) \pi = \int_{-\infty}^{z_0}dz~\sqrt {{\lambda e^{2z}\over \left(\left(1+e^{z}\right)^{3/2}-1\right)~\sqrt{1+e^{z}}} } = 2\sqrt \lambda \int_{w=1}^{\sqrt{1+e^{z_0}}}~{dw\over\sqrt{w^3-1}}$$
$$\approx  2\sqrt \lambda\left[{2\sqrt\pi\Gamma\left(7/6\right)\over \Gamma\left(2/3\right)} - 2\sqrt {{16\lambda\over9} + \left({16\lambda\over9}\right)^3{B^2\over8}}~\right] ,\hspace{1cm}\eqno{(2.35)}$$
which implies the following spectrum of the $0^{++}$  glueball:
$$M =\left(n+{1\over 2}\right)\pi^2T\left[{\Gamma\left(2/3\right)\over 2\sqrt\pi\Gamma\left(7/6\right)}\right] \left(1- {B^2\over6}{\Gamma\left(2/3\right)\over 2\sqrt\pi\Gamma\left(7/6\right)}\left({{\left(n+{1\over2}\right)}\pi\over2}\right)^{3/2} \right).\eqno{(2.36)}$$
Therefore we  see that the dipole has an effect to produce attractive force between the gluons within the glueball.
 \subsubsection{Spectrum of  $1^{--}$  Glueball}
Consider next the $1^{--}$  glueball mass.  In the supergravity description we have to solve the wave equation of antisymmetric tensor field $A_{\mu\nu}$ in (2.15).  As before we  look for $\theta_i$-independent solution of the form 
$$A_{\mu\nu} = \rho_{\mu\nu}(U) e^{ik\cdot x},~~~~~ k_\mu=\left(k_0,0,0,0\right). \eqno{(2.37)}$$
In searching the $1^{--}$  glueball spectrum we let $\rho_{\mu\nu}(U) = \rho(U) \delta_{\mu1}\delta_{\nu2}$ [10] and the equation for $\rho$ becomes
$$\partial_U\left({U^{-1/2}\over \sqrt{1+B^2U^2}}~\left(U^3-U_T^3\right)\partial_U \rho \right) +{U^{-1/2}\over  \sqrt{1+B^2U^2}}\left({M^2}\left(1-\gamma B^2U^2\right) - {16\over3}U^{3\over2}\right) \rho  = 0. \eqno{(2.38)}$$
As before we let $U=\sqrt y$, $ y= y_T(1+e^z)$  and define $\Psi(z) = \sqrt{f(z)} ~\rho(z)$, then the above equation becomes 
$$ \Psi''(z) + V(z) \Psi(z)=0,~~~~V(z) =V_0(z) -{1\over2}{f''(z)\over f(z)}+{1\over4}\left({f'(z)\over f(z)}\right)^2. \eqno{(2.39)}$$
where
$$V_0(z) ={M^2\over 4\sqrt {y_T}}{e^{2z}\over \left(\left(1+e^{z}\right)^{3/2}-1\right)~\left(1+e^{z}\right)} - {4\over3}{y_T^{1/4}~e^{2z}\over \left(\left(1+e^{z}\right)^{3/2}-1\right)~\left(1+e^{z}\right)^{1/4}}\eqno{(3.40)}$$
 $$f(z) = {\left(1+e^{z}\right)^{1/4}e^{-z}\over\sqrt{1+B^2y_T\left(1+e^{z}\right)}}\left(\left(1+e^{z}\right)^{3/2}-1\right).\hspace{5cm}\eqno{(2.41)}$$
As before, in the case of $M \gg 1$ we can determine the two turning points and the mass spectrum of the $1^{--}$  glueball evaluated from the WKB approximation becomes
$$M =\left(n+{1\over 2}\right)\pi^2T\left[{\Gamma\left(2/3\right)\over 2\sqrt\pi\Gamma\left(7/6\right)}\right] \left(1- {28B^2\over9}{\Gamma\left(2/3\right)\over 2\sqrt\pi\Gamma\left(7/6\right)}\left({{\left(n+{1\over2}\right)}\pi\over2}\right)^{3/2} \right).\eqno{(2.42)}$$
Therefore we  see that the dipole has an effect to produce attractive force between the gluons within the glueball.
\section{Baryon in Dipole Field Theory}
The dual baryon configuration suggested in [7,8] contains two contributions.  The first is the string stretched between the boundary of the $AdS_5$ space and the second  is the D5-brane wrapped on the $S^5$.  They are of the same order and  we should consider both of them.

The action for a static D5-brane wrapped on dipole deformed $S^5$ in (2.1)  is 
$$S_{D_5} = {1\over (2\pi)^5}\int dx^6 e^{-\phi}\sqrt g = {TNU_0\over 8\pi} + O(B^4),\eqno{(3.1)}$$
in which we neglect the terms higher then $B^2$ order.  Note that  $T$ is the time period and $U_0$ is the location of the baryon vertex in the bulk.

The $N$ strings we considered is such that the strings end on boundary of the $AdS_5$ space with radius $L$ in a symmetric way.  The string may be static or moving under the NS-NS B field, which are investigated in the follow.

\subsection{Static String Configuration}
The static string described by the Nambu-Goto action in the gauge 
$$t=\tau,~~~~~z=\sigma,~~~~~~U=U(z), \eqno{(3.2)}$$
 is
$$S_{F_1} = {T\over 2\pi}\int dz~\sqrt{\left(\partial_z U\right)^2 + {U^4\over R^4 \left(1+ B^2 U^2\right)}}.\eqno{(3.3)}$$
The total action is the summation of (3.1) and (3.3) and the variation of it under $U\rightarrow U+\delta U$  contains a volume term as well as a surface term. 
The volume term leads to 
$${ {U^4\over \left(1+ B^2 U^2\right)} \over\sqrt{\left(\partial_z U\right)^2 + {U^4\over R^4 \left(1+ B^2 U^2\right)}}} =constant, \eqno{(3.4)}$$
because the Lagrangian of  action $S_{F_1}$ does not depend explicitly on z.  The surface term yields the relation
$$\delta U {TN\over 8\pi} =\delta U{TN (\partial U_0)\over 2\pi\sqrt{\left(\partial_z U_0\right)^2 + {U_0^4\over R^4 \left(1+ B^2 U_0^2\right)}}},\eqno{(3.5)}$$
which implies that
$$ (\partial_z U_0)^2 = {U_0^4\over 15R^4 \left(1+ B^2 U_0^2\right)}.\eqno{(3.6)}$$
Using this relation and consider the constant in (3.4) as the value at $U_0$ we find that 
$$(\partial_zU)^2 = {U^4\over R^4 \left(1+ B^2 U^2\right)}\left[{16\over15}{\left(1+ B^2 U^2_0\right)U^4\over \left(1+ B^2 U^2\right)U_0^4}-1\right].\eqno{(3.7)}$$
With the help of this relation the radius of the baryon could be calculated by  
$$ L = \int dz = \int {dU \over \partial_z U} = \int_{U_0}^\infty dU~{R^2 \sqrt{1+ B^2 U^2}\over U^2\sqrt{{16\over15}{\left(1+ B^2 U^2_0\right)U^4\over \left(1+ B^2 U^2\right)U_0^4}-1}}={R^2\over U_0} \int_{1}^\infty dy~{\sqrt{1+ B^2U_0^2 y^2}\over y^2\sqrt{{16\over15}{\left(1+ B^2U_0^2 \right)y^4\over 1+ B^2 U_0^2 y^2}-1}}$$
$$
= {R^2\over U_0} \int_0^1 dx~{{x^2+ B^2U_0^2}\over \sqrt{{16\over15}\left(1+ B^2U_0^2 \right) -x^2 \left(x^2+B^2 U_0^2\right)}}.\eqno{(3.8)}$$
Substituting the relation (3.7) into (3.3) the part of the baryon energy that comes from the N string could be evaluated by the formula
$$M_{F1} = {U_0\over 2\pi} \int_{1}^\infty dy~\left({\sqrt{16\over15}\sqrt{1+ B^2U_0^2 }~y^2\over \sqrt{{16\over15}{\left(1+ B^2U_0^2 \right)y^4- (1+ B^2 U_0^2 y^2)}}}-1\right) - {U_0\over 2\pi}.\eqno{(3.9)}$$
We can therefore use the equations (3.8) and (3.9) to find the value of  $M_{F1}$ which becomes the function of baryon radius $L$.

\subsubsection{Small Baryon : Minimum Radius}
A numerical evaluation of (3.8) is presented in figure 1.
\\
\\
\scalebox{1}{\hspace{5cm}\includegraphics{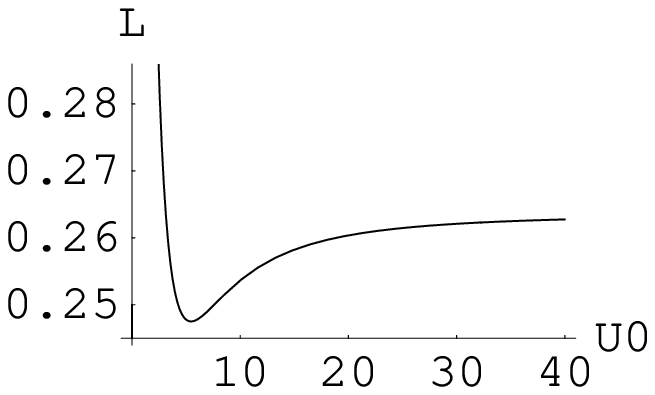}}
\\
\\
{\hspace{1cm} {\it Figure 1.  The function $L(U_0)$ at $B=0.2$.    We see that baryon has  a minimum radius.}
\\
\\
Thus we see that the baryon with a dipole will has a minimum radius.  The existence of the minimum radius could also be directly analyzed from (3.8). 

In the limit of $U_0\rightarrow \infty$  eq.(3.8) gives the following approximation
$$ L(U_0\rightarrow\infty) \approx {R^2\over U_0} \int_0^1 dx~{B^2U_0^2\over B U_0\sqrt{{16\over15} -x^2 }} = R^2 B \sin^{-1}\left(\sqrt{16\over15}\right).\eqno{(3.10)}$$
Thus the minimum radius is proportional to the dipole length $B$ and $\sqrt N$ (note that $R^4=4\pi N$).  However, it shall be noticed that, form figure 1 we see that the real minimum value of the baryon radius is less then  above value.  
\subsubsection{Big Baryon }
The big baryon with large radius $L$ is that with small $U_0$.  At small value of  $U_0$ the equations (3.8) and (3.9) could be approximated as  
$$ L \approx  {R^2\over U_0}\left[ \int_0^1 dx~{x^2\over \sqrt{{16\over15}-x^4}}+B^2U_0^2 \int_0^1 dx~\left[{1\over \sqrt{{16\over15}-x^4}}-{1\over2}{x^2({16\over15}-x^2)\over \left({16\over15}-x^4\right)^{3/2}}~\right]\right]\hspace{2cm}$$
$$= {R^2\over U_0}\left[0.481 + 0.994 B^2 U_0^2\right].\hspace{8cm}\eqno{(3.11)}$$
$$M_{F1} \approx {U_0\over 2\pi} \int_{1}^\infty dy~\left[{\sqrt{16\over15}y^2\over \sqrt{{16\over15}y^4- 1}}+
B^2U_0^2 \sqrt{16\over15} {y^2\over2} \left({1\over \sqrt{{16\over15}y^4- 1}} -{{16\over15}y^4-y^2\over \left({16\over15}y^4-1\right)^{3/2}}\right)-1\right] - {U_0\over 2\pi}$$
$$ ={U_0\over 2\pi}\left[ 0.284 + 0.398 B^2U_0^2\right] - {U_0\over 2\pi}.\hspace{7cm}\eqno{(3.12)}$$
\\
We can use  (3.11) to express  $U_0$ as function $L$
$$ U_0 = {R^2\over L}\left[ 0.481 + 0.230 B^2{R^4\over L^2}\right].  \eqno{(3.13)}$$
Substituting above relation into (3.12) and (3.1) we finally find the baryon energy  
$$H \equiv M_{F1}+M_{D_5} \approx  {1\over 2\pi} \left(-{0.224 R^2\over L} -{0.063 R^6\over L^3} B^2 \right),\eqno{(3.14)}$$
in which $M_{D_5}$ is the energy contribution from D5 brane in (3.1).  Above results tell us  that dipole could produce an attractive force between the quarks to reduce the baryon energy.   

As mentioned in section II, the string  in a background with  $B$ field is somewhat analogous to the situation when a charged particle enters a region with a magnetic field.  Thus, the string may be moving with  a velocity.   So let us consider the case with moving string in below.

\subsection{Moving String Configuration}
The moving string we considered is described by the gauge [20]
$$t=\tau,~~~~~\theta_3=\omega t,~~~~~z=\sigma,~~~~~~U=U(z). \eqno{(3.15)}$$
The Nambu-Goto action in this gauge is
$$S_{F_1} = {T\over 2\pi}\int dz~\sqrt{\left(1-{R^4\over U^2}{\omega^2\over 1+B^2U^2}\right)\left(\left(\partial_z U\right)^2 + {U^4\over R^4 \left(1+ B^2 U^2\right)}\right)} - {B\omega U^2\over 1+B^2U^2}.\eqno{(3.16)}$$
As before, the variation of  volume term gives 
 $${U^4\over R^4 \left(1+ B^2 U^2\right)}{\sqrt{1-{R^4\over U^2}{\omega^2\over 1+B^2U^2}}\over \sqrt{\left(\partial_z U\right)^2 + {U^4\over R^4 \left(1+ B^2 U^2\right)}}} +  {B\omega U^2\over 1+B^2U^2} =  constant,\eqno{(3.17)}$$
because the Lagrangian of  action $S_{F_1}$ does not depend explicitly on z.  The variation of  surface term gives 
$$ (\partial_z U_0)^2 = {U_0^4\over R^4 \left(15 \left(1+ B^2 U_0^2\right) - {16R^4 \omega^2\over U_0^2 }\right)}.\eqno{(3.18)}$$
Using this relation and consider the constant in (3.4) as the value at $U_0$ we have a relation
$$\left(\partial_z U\right)^2 = {{U^8\over R^8 \left(1+ B^2 U^2\right)^2}\left(1-{R^4\over U^2}{\omega^2\over 1+B^2U^2}\right)\over \left({\sqrt{15 \left(1+ B^2 U_0^2\right) - {16R^4 \omega^2\over U_0^2 }}\over4}{U_0^2\over R^2 \left(1+ B^2 U_0^2\right)}  + {B\omega U_0^2\over 1+B^2U_0^2} - {B\omega U^2\over 1+B^2U^2} \right)^2}-  {U^4\over R^4 \left(1+ B^2 U^2\right)}.\eqno{(3.19)}$$
Using above relation we will determine the angular velocity $\omega$.

In the limit $U\rightarrow \infty$ above relation becomes
$$\left(\partial_z U\right)^2 \rightarrow {{U^4\over R^8 B^4}\over \left[{\sqrt{15 \left(1+ B^2 U_0^2\right) - {16R^4 \omega^2\over U_0^2 }}\over4}{U_0^2\over R^2 \left(1+ B^2 U_0^2\right)}  + {B\omega U_0^2\over 1+B^2U_0^2} - {\omega \over B} \right]^2}.\eqno{(3.20)}$$
Thus, if the bracket term in above equation is zero then, for the fixed values of $U_0$ and $B$ there will have a minimum radius $L$ as can be seen from the relation $ L = \int dz = \int {dU \over \partial_z U}$ which has been used in (3.8).  While the dipole has effect to produce attractive force the baryon with small radius will therefore has less energy. Thus the least energy of dual moving string  will rotate with an angular velocity which is the solution by letting the bracket term in (3.20) to be zero.  This property give a very simple relation between angular velocity $\omega$ and dipole length $B$ :
$$ \omega^2= {15 B^2\over 16}{U_0^4\over R^4}\eqno{(3.21)}$$
Substituting this relation into (3.19) we also find a very simple relation 
$$(\partial_zU)^2 = {U^4\over R^4 }\left[{16\over15}{U^4\over U_0^4}-1\right].\eqno{(3.22)}$$
This relation is just (3.7) while let $B=0$. Thus we find that a moving string has the same result as that without dipole field.  This seems a  surprise property at first sight.  The reason behind it may be argued  as following.  

The dual string  in a background with  $B_{z \theta_3}$ field is somewhat analogous to the situation when a charged particle enters a region with a magnetic field.  Thus, the string will be rotating along $\theta_3$ with a constant angular momentum $\omega$ which is proportional to the strength of the NS-NS field, as shown in (3.21).   The configuration described in (3.15) has a binding energy  (it is negative) from B field which will be  just  canceled by the kinetic energy (it is positive) from the moving.  Thus the moving dual string does not depend on the  value of dipole field and we have the same result as that without dipole field.

For clear we plot in figure 2 the baryon energy calculated from static and moving strings.
\\
\\
\scalebox{1}{\hspace{4cm}\includegraphics{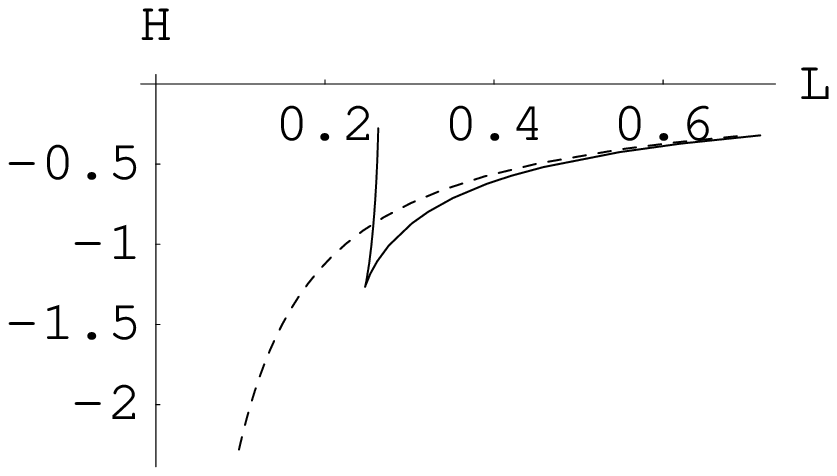}}
\\
\\
{\hspace{0cm} {\it Figure 2.  The baryon energy calculated from static string (in solid line) and moving string (in dashed line) at $B=0.2$.   We see that a small baryon could be found in the dual moving string which then transit to the static string configuration of big baryon.}
\\
\\
In conclusion, as the static string configuration shows an attractive force the baryon will therefore be in the static string configuration which, however could exist only if it has a radius larger then a critical value.   A small baryon could be  found in the dual moving string which is that transit from the static configuration as shown in figure 2.  The transition is first order as the energy of the two configurations at transition radius is discontinuous.
\\
\section{Conclusion}
 In this paper, we first follow [10] to investigate the glueball spectrum in the supersymmetric and non-supersymmetric 4D non-commutative dipole gauge theory from the  holographic description. To find the analytic formula of the spectrum of $0^{++}$ and $1^{--}$ glueball we adopt the semiclassical WKB approximation to solve the dilaton and antisymmetric tensor field equations on the dual supergravity backgrounds, respectively.  In the supersymmetric theory we see that the dipole length plays the intrinsic scale which reflects the discrete spectrum therein.  In the non-supersymmetric theory,  the temperature (or the radius of compactification)  in there will now play the intrinsic scale and  we see that the dipole has an effect to produce attractive force between the gluons within the glueball. 

 We next study the baryon energy following the method in [8].  To study the confining force between the quarks within the baryon we consider the strings that ``hang" into the dipole deformed AdS geometry and regarded the baron vertex as the strings and fivebrane which are described in terms of separated actions. We first consider the string as a static configuration and find that the baryon could only exist if it is larger then a critical radius.   We see that the dipole could also produce an attractive force between the quarks.  We next consider the string as a configuration moving with an angular velocity.  We find that the angular velocity is proportional to the NS-NS B field.  After the evaluation we find that the baryon has two phases in which a big baryon is dual to the static string while a small baryon is described by a moving dual  string.    The phase transition property is like that in our study of the Wilson loop of non-commutative gauge theory  form dual string description [20]. 

 Finally, the baryon  vertex  constructed by Callen et.al. [9] is considered as the D5 brane wrapped on an $S^5$ on which N fundamental strings terminate and they are dissolved in it [7].   While the method could be used to investigate the more details of the baryon vertex the mathematics therein is quite involved.  It is interesting to see how the property of phase transition from small baryon to big baryon would be shown in the Callen method. We will investigate this problem in the next  paper.  The properties of  glueball and baryon on the other dipole field deformed background  [21] are of interesting and remain to be  studied.

\newpage
\begin{center} {\bf REFERENCES}\end{center}
\begin{enumerate}
\item J.~M.~Maldacena, ``The large N limit of superconformal field theories and supergravity,'' Adv. Theor. Math.  Phys. 2 (1998) 231 [hep-th/9711200];
 S. S. Gubser, I. R. Klebanov and A.~M.~Polyakov, ``Gauge theory correlators from non-critical string theory,'' Phys. Lett. B428 (1998) 105 [hep-th/9802109]; E.~Witten, ``Anti-de Sitter space and holography,'' Adv. Theor. Math. Phys. 2 (1998) 253 [hep-th/9802150].
\item E.~Witten, ``Anti-de Sitter space, thermal phase transition, and confinement in  gauge theories,'' Adv. Theor. Math. Phys. 2 (1998) 505 [hep-th/9803131].
\item J. M. Maldacena, ``Wilson loops in large N field theories", Phys. Rev. Lett. 80 (1998) 4859 [hep-th/9803002]; S. J. Rey and J. Yee, Macroscopic strings as heavy quarks in large N theory and anti-de Sitter supergravity", Eur. Phys. J. C22 (2001) 379 [hep-th/9803001].
\item N.~Drukker and B.~Fiol, ``All-genus calculation of Wilson loops using
  D-branes,''  JHEP  02 (2005) 010 [hep-th/0501109]; S.~A. Hartnoll and S.~Prem~Kumar,  ``Multiply wound Polyakov loops at strong coupling,''  Phys. Rev.  D74 (2006) 026001[hep-th/0603190]; S.~Yamaguchi, ``Wilson loops of anti-symmetric representation and  D5-branes,''  JHEP 05 (2006) 037 [hep-th/0603208].
\item  A. Karch and E. Katz, ``Adding avor to AdS/CFT", JHEP 06 (2002) 043,
[hep-th/0205236];  A. Karch, E. Katz, and N. Weiner, ``Hadron masses and screening from AdS Wilson loops", Phys. Rev. Lett. 90 (2003) 091601 [hep-th/0211107];  M. Kruczenski, D. Mateos, R. C. Myers, and D. J. Winters, ``Meson spectroscopy in AdS/CFT with  flavour", JHEP 07 (2003) 049 [hep-th/0304032]; J. Babington, J. Erdmenger, N. J. Evans, Z. Guralnik, and I. Kirsch, ``Chiral symmetry breaking and pions in non-supersymmetric gauge / gravity duals", Phys. Rev. D69 (2004) 066007 [hep-th/0306018].
\item   J. Erdmenger, N. Evans, I. Kirsch, and E. Threlfall, `` Mesons in Gauge/Gravity Duals - A Review,''  arXiv:0711.4467 [hep/th].
\item E. Witten, ``Baryons and Branes in Anti de Sitter Space," J. High Energy Phys.  07 (1998) 006  [hep-th/9805112];  H. Ooguri and D. Gross, ``Aspects of Large N Gauge Theory Dynamics as seen by String theory," Phys. Rev. D58 (1998) 106002 [hep-th/9805129]. 
\item  A. Brandhuber, N. Itzhaki, J. Sonnenschein, and S. Yankielowicz, ``Baryons from Supergravity," JHEP 07 (1998) 020 [hep-th/9806158]; Y. Imamura, ``Mass and Phase Transitions in Large N Gauge Theory," Prog. Theor. Phys. 100 (1998) 1263 [hep-th/9806162].
\item Y. Imamura, ``Supersymmetries and BPS Congurations on Anti-de Sitter Space,"  Nucl. Phys. B537 (1999) 184 [hep-th/9807179] ;  C. G. Callan, A. Guijosa, and K. Savvidy, ``Baryons and String Creation from the Fivebrane Worldvolume Action," Nucl. Phys. B547 (1999) 127 [hep-th/9810092];  C. G. Callan, A. Guijosa, K. Savvidy, and O. Tafjord, ``Baryons and Flux Tubes in Confining Gauge Theories from Brane Actions", Nucl.Phys. B555 (1999) 183  [hep-th/9902197].
\item C. Csaki, H. Ooguri, Y. Oz and J. Terning,``Glueball mass spectrum from supergravity," JHEP 9901 (1999) 017 [hep-th/9806021]; C. Csaki, Y. Oz, J. Russo and J. Terning, ``Large N QCD from rotating branes," Phys. Rev. D 59 (1999) 065012 [hep-th/9810186]; R. de Mello Koch, A. Jevicki, M. Mihailescu and J. P. Nunes, ``Evaluation of glueball masses from supergravity," Phys. Rev. D 58 (1998) 105009 [hep-th/9806125]; J. A. Minahan, ``Glueball mass spectra and other issues for supergravity duals of QCD models," JHEP 9901 (1999) 020 [hep-th/9811156] 
\item R. C. Brower, S. D. Mathur and C. I. Tan, ``Glueball spectrum for QCD from AdS supergravity duality," Nucl. Phys. B 587 (2000) 249 [hep-th/0003115].
\item H. Boschi-Filho and N. R. F. Braga. ``Gauge/string duality and scalar glueball mass ratios", JHEP 0305 (2003) 009 [hep-th/0212207];
H. Boschi-Filho, N. R. F. Braga, H. L. Carrion,``Glueball Regge trajectories from gauge/string duality and the Pomeron", Phys.Rev. D73 (2006) 047901 [hep-th/0507063];  R. Apreda, D. E. Crooks, N. Evans, M. Petrini  ``Confinement, Glueballs and Strings from Deformed AdS", JHEP 0405 (2004) 065 [hep-th/0308006].
\item T. Nakajima, K. Suzuki, H. Takahashi, `` Glueball mass spectra for supergravity duals of noncommutative gauge theories", JHEP 0601 (2006) 016 [hep-th/0508054].
\item P. Colangelo, F. De Fazio, F. Jugeau, S. Nicotri. ``On the light glueball spectrum in a holographic description of QCD ", [hep-ph/0703316];  P.Colangelo, F.De Fazio, F.Jugeau, S. Nicotri, ``Investigating AdS/QCD duality through scalar glueball correlators", arXiv: 0711.4747 [hep-th].
\item A. Bergman and O. J. Ganor,``Dipoles, Twists and Noncommutative Gauge Theory," JHEP 0010 (2000) 018 [hep-th/0008030]; A. Bergman, K. Dasgupta, O. J. Ganor, J. L. Karczmarek, and G. Rajesh,``Nonlocal Field Theories and their Gravity Duals," Phys.Rev. D65 (2002) 066005 [hep-th/0103090].
\item  M. Alishahiha and H. Yavartanoo,``Supergravity Description of the Large N Noncommutative Dipole Field Theories," JHEP 0204 (2002) 031 [hep-th/0202131] . 
\item Wung-Hong Huang, ``Wilson-t'Hooft Loops in Finite-Temperature Non-commutative Dipole Field Theory from Dual Supergravity,''   Phys.Rev. D76 (2007) 106005, arXiv: 0706.3663 [hep-th].
\item K. Dasgupta and M. M. Sheikh-Jabbari, ``Noncommutative Dipole Field Theories," JHEP 0202 (2002) 002 [hep-th/0112064]; N. Sadooghi and M. Soroush, ``Noncommutative  Dipole QED", Int. J. Mod. Phys. A18 (2003) 97 [hep-th/0206009]. 
\item J. M. Maldacena and J. G. Russo,`` Large N Limit of Non-Commutative Gauge Theories," JHEP 9909 (1999) 025 [hep-th/9908134]; A. Hashimoto and N. Itzhaki,``Non-Commutative Yang-Mills and the AdS/CFT Correspondence," Phys.Lett. B465 (1999) 142 [hep-th/9907166]. 
\item Wung-Hong Huang, ``Dual String Description of Wilson Loop in Non-commutative Gauge Theory ",  Phys.Lett.B647 (2007) 519; Erratum-ibid.B652 (2007) 388 [hep-th/0701069].
\item U. Gursoy and C. Nunez   ,  ``Dipole Deformations of N=1 SYM and Supergravity backgrounds with U(1) X U(1) global symmetry ,'' Nucl.Phys. B725 (2005) 45-92  [hep-th/0505100 ]; N.P. Bobev, H. Dimov, R.C. Rashkov ,  ``Semiclassical Strings, Dipole Deformations of N=1 SYM and Decoupling of KK Modes,'' JHEP 0602 (2006) 064 [hep-th/0511216].
\end{enumerate}
\end{document}